\documentclass[10pt,aps,prb,superscriptaddress,twocolumn,floatfix]{revtex4-2}
\usepackage{graphicx}
\usepackage{dcolumn}
\usepackage{bm}
\usepackage{amsmath}
\usepackage{subfigure}
\usepackage{xfrac}
\usepackage{hyperref}
\usepackage{url}
\usepackage[T1]{fontenc}

\begin{document}

\title{Topological melting of the metastable skyrmion lattice in the chiral magnet Co$_9$Zn$_9$Mn$_2$}

\author{Victor Ukleev}
\email{victor.ukleev@psi.ch}
\affiliation{Laboratory for Neutron Scattering and Imaging (LNS), Paul Scherrer Institute (PSI), CH-5232 Villigen, Switzerland}
\affiliation{Helmholtz-Zentrum Berlin f\"ur Materialien und Energie, D-14109 Berlin, Germany}
\author{Daisuke Morikawa}
\affiliation{RIKEN Center for Emergent Matter Science (CEMS), Wako 351-0198, Japan}
\affiliation{Institute of Multidisciplinary Research for Advanced Materials (IMRAM), Tohoku University, Sendai 980-8577, Japan}
\author{Kosuke Karube}
\affiliation{RIKEN Center for Emergent Matter Science (CEMS), Wako 351-0198, Japan}
\author{Akiko Kikkawa}
\affiliation{RIKEN Center for Emergent Matter Science (CEMS), Wako 351-0198, Japan}
\author{Kiyou Shibata}
\affiliation{RIKEN Center for Emergent Matter Science (CEMS), Wako 351-0198, Japan}
\affiliation{Institute of Industrial Science, The University of Tokyo, Tokyo 153–8505, Japan}
\author{Yasujiro Taguchi}
\affiliation{RIKEN Center for Emergent Matter Science (CEMS), Wako 351-0198, Japan}
\author{Yoshinori Tokura}
\affiliation{RIKEN Center for Emergent Matter Science (CEMS), Wako 351-0198, Japan}
\affiliation{Department of Applied Physics, University of Tokyo, Tokyo 113-8656, Japan}
\affiliation{Tokyo College, University of Tokyo, Tokyo 113-8656, Japan}
\author{Taka-hisa Arima}
\affiliation{RIKEN Center for Emergent Matter Science (CEMS), Wako 351-0198, Japan}
\affiliation{Department of Advanced Materials Science, University of Tokyo, Kashiwa 277-8561, Japan}
\author{Jonathan S. White}
\affiliation{Laboratory for Neutron Scattering and Imaging (LNS), Paul Scherrer Institute (PSI), CH-5232 Villigen, Switzerland}



\begin{abstract}

In a $\beta$-Mn-type chiral magnet Co$_9$Zn$_9$Mn$_2$, we demonstrate that the magnetic field-driven collapse of a room temperature metastable topological skyrmion lattice passes through a regime described by a partial topological charge inversion. Using Lorentz transmission electron microscopy, the magnetization distribution was observed directly as the magnetic field was swept antiparallel to the original skyrmion core magnetization, i.e. negative magnetic fields. Due to the topological stability of skyrmions, a direct transition of the metastable skyrmion lattice to the equilibrium helical state is avoided for increasingly negative fields. Instead, the metastable skyrmion lattice gradually transforms into giant magnetic bubbles separated by $2\pi$ domain walls. Eventually these large structures give way to form a near-homogeneously magnetized medium that unexpectedly hosts a low density of isolated skyrmions with inverted core magnetization, and thus a total topological charge of reduced size and opposite sign compared with the initial state. A similar phenomenon has been observed previously in systems hosting ordered lattices of magnetic bubbles stabilized by the dipolar interaction and called “topological melting”. With support from numerical calculations, we argue that the observed regime of partial topological charge inversion has its origin in the topological protection of the starting metastable skyrmion state.

\end{abstract}
\maketitle

\section{Introduction}

A magnetic skyrmion is a topological object with vortex-like swirling spin configuration, which is characterized by an integer termed the topological number:

\begin{equation}
n={\tfrac{1}{4\pi}}\int\mathbf{M}\cdot\left(\frac{\partial \mathbf{M}}{\partial x}\times\frac{\partial \mathbf{M}}{\partial y}\right)dx dy.
\label{eq1}
\end{equation}
Here $\mathbf{M}$ is a unit vector describing the direction of a smoothly varying local magnetization. A skyrmion crystal (SkX) is a periodic, usually triangular-coordinated arrangement of skyrmion. It is a non-trivial collective topological magnetic state that is robust against perturbations, such as magnetic fields and thermal agitation, since it cannot be continuously transformed into an alternative state with smoothly varying magnetization which is described by a different topology \cite{bogdanov1994thermodynamically,nagaosa2013topological}. Indeed, in real systems the stability of skyrmions is limited due to the violation of the continuous limit at the atomic scale, the presence of defects or edges, and temperature fluctuations. In the past decade, several kinds of magnets have been found to host skyrmion ($n=-1$) \cite{muhlbauer2009skyrmion,yu2011near,seki2012observation,tokunaga2015new,kezsmarki2015neel,kurumaji2017neel}, bi-skyrmion ($n=-2$) \cite{yu2014biskyrmion,lee2016synthesizing,wang2016centrosymmetric,takagi2018low}, or anti-skyrmion ($n=+1$) \cite{nayak2017magnetic} textures. 
Note that since $n$ reverses its sign under time reversal, the above topological numbers are given for a common polarity of the core \cite{koshibae2016theory}. \\

Skyrmions can be manipulated by ultra-low density electric currents, making them promising for spintronics applications \cite{nagaosa2013topological,fert2017magnetic}. In bulk materials, skyrmions have been studied extensively in chiral cubic magnets such as the $B20$-type compounds \cite{muhlbauer2009skyrmion,yu2011near,adams2012long,munzer2010skyrmion,kanazawa2012possible,seki2012observation} and $\beta-$Mn-type Co-Zn-Mn alloys \cite{tokunaga2015new,karube2016robust,karube2017skyrmion,morikawa2017deformation,karube2018sciadv,ukleev2019element,nagase2021observation}. In these compounds, due to the interplay between the exchange interaction and the Dzyaloshinskii-Moriya interaction (DMI), a helical magnetic ordering becomes stable upon cooling below a critical temperature ($T_c$) in zero field. By the application of a moderate magnetic field just below $T_c$, the helical structure transforms into the ordered triangular SkX. At the mean-field level \cite{bak1980theory}, both the characteristic helical pitch $\lambda$ and skyrmion size are determined by the ratio of the exchange stiffness $A_{ex}$ to the Dzyaloshinskii constant $D$, according to $\lambda=4\pi A_{ex}/D$.\\

\begin{figure*}
\begin{center}
\includegraphics[width=1\linewidth]{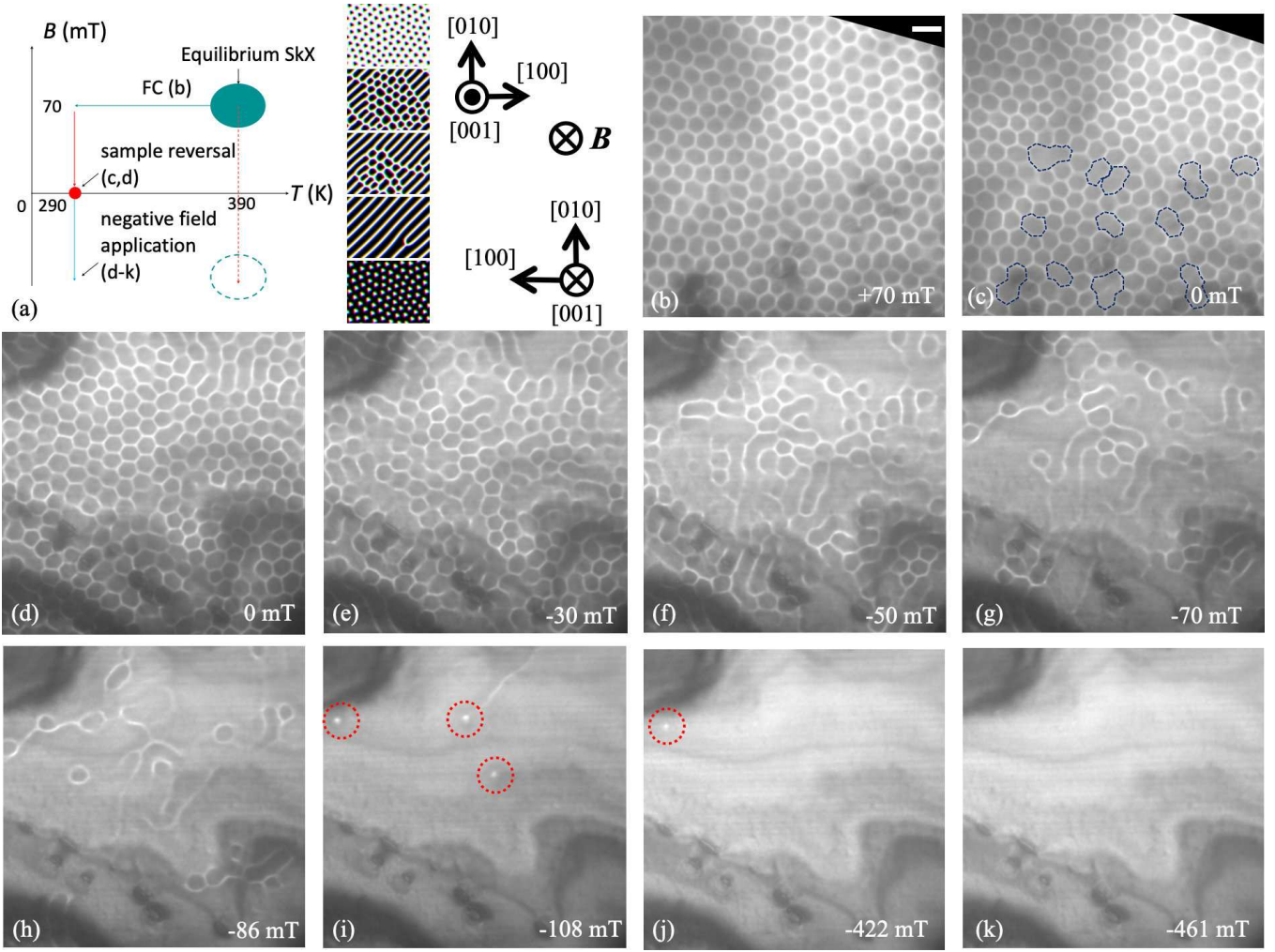}
\caption{(a) Schematic illustration of the measurement process. On the left side of (a), the labels (b-k) describes the points in the schematic phase diagram where the data shown in the corresponding panels were collected. The middle of (a) show a sketch the field-driven magnetic texture reversal of the equilibrium phases just below $T_c$ \cite{kovacs2016lorentz}. The right side of (a) shows the crystallographic orientation of the thin plate with respect to the applied magnetic field before (top panel) and after sample reversal (bottom panel), respectively (see text for details). (b) Over-focused LTEM images on the (001) plane after the FC process from 390\,K to 290\,K at 70\,mT. The crystal orientation and field direction are indicated in (a). (c) LTEM image of the zero-field SkX. Merged skyrmions are indicated by dashed blue lines. (d -- k) Under-focused LTEM images of the sample after reversal. Each panel corresponds to the evolution of the magnetic structure upon application of the magnetic field in the negative direction compared to the original state. Skyrmions with reversed cores are indicated in panels (i -- j) by dashed red circles.The scale bar shown in (b) is 200\,nm.}
\label{fig1}
\end{center}
\end{figure*}

\begin{figure*}
\begin{center}
\includegraphics[width=1\linewidth]{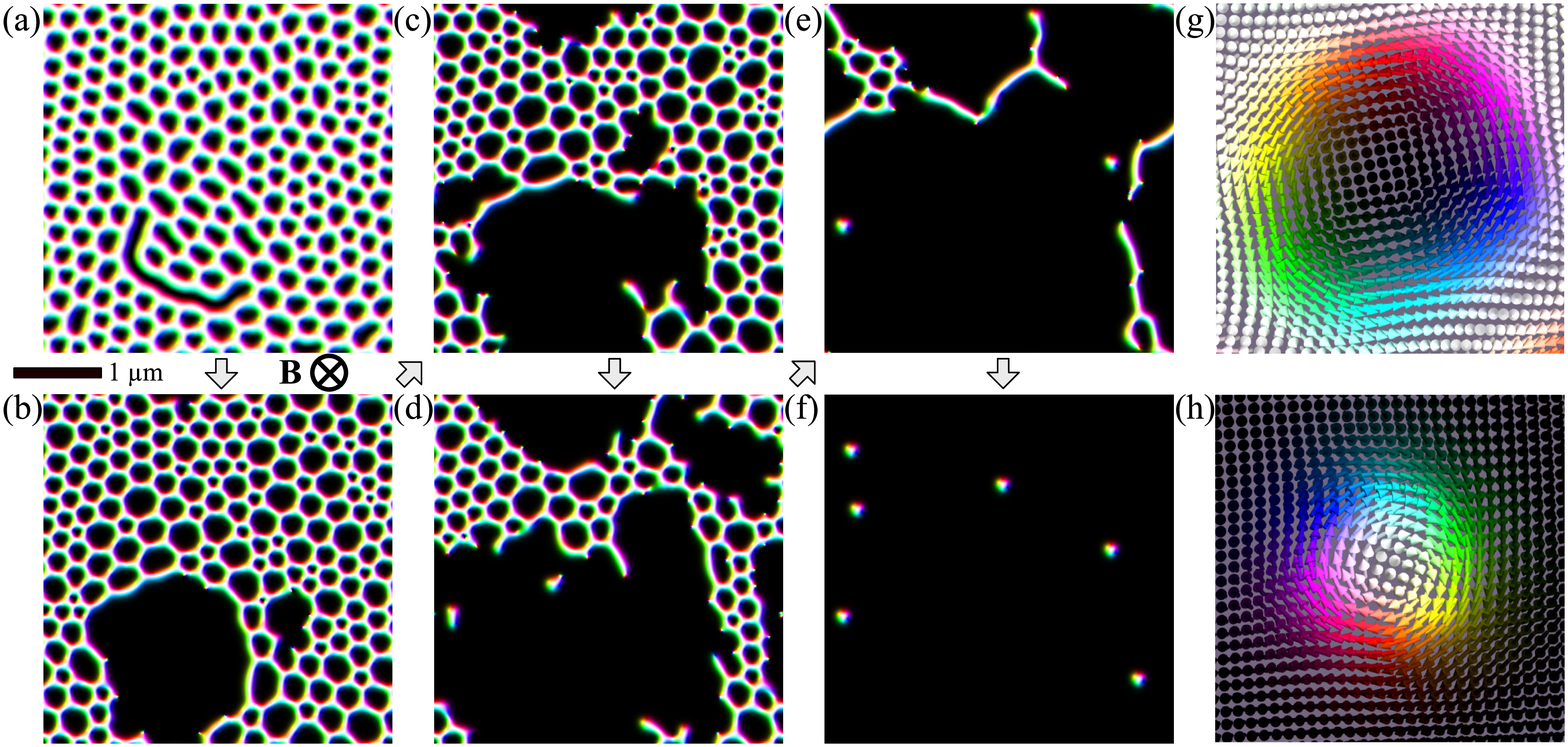}
\caption{Micromagnetic simulations of the SkX in a Co$_9$Zn$_9$Mn$_2$ thin plate in (a) zero and negative magnetic fields (b) $B=-280$\,mT, (c) $B=-285$\,mT, (d) $B=-290$\,mT, (e) $B=-295$\,mT, and (f) $B=-310$\,mT. Detailed illustrations of the magnetization distributions for the original zero-field and inverted skyrmions shown in panels (a--f) are given in panels (g,h), respectively. The in-plane magnetization direction is indicated by the colors shown in panels (g,h).}
\label{fig2}
\end{center}
\end{figure*}

In contrast, periodic lattices of magnetic bubbles can be stabilized in thin films of centrosymmetric magnets with easy axis anisotropy \cite{thiele1969theory,malozemoff1975high,bogatyrev2018makes}. The in-plane rotation sense of the magnetization around the bubble core is non-degenerate, thus leading to a cancellation of the total magnetic chirality and therefore the topological charge of the bubble lattice \cite{yu2017variation,loudon2019images}. This is different from skyrmion systems where the magnetic chirality is determined by the sign of DMI which itself is determined by the handedness  either the chiral crystal, \cite{grigoriev2009crystal,morikawa2013crystal} or the interfacial symmetry breaking in magnetic multilayer systems. The research of magnetic bubble domains has existed since the second half of the 20th century, leading to many intriguing observations and theoretical developments. In particular, a coherent inversion of the bubble lattice in garnet films has been experimentally observed upon reversing the direction of applied magnetic field and called "topological switching" \cite{o1973dynamics,papworth1974topological,malozemoff1975high,randoshkin1978dynamics}. An alternative mechanism of magnetic bubble lattice collapse via the formation of cellular domains was called "topological melting" \cite{babcock1989topological,zablotskii1995ordering}. Both scenarios were successfully reproduced numerically via a fine-tuning of short-range ferromagnetic and long-range antiferromagnetic dipolar interactions \cite{jagla2004numerical}. According to a more recent theory that includes the DMI, it is suggested that the magnetization switching processes of both magnetic skyrmions in chiral magnets, \cite{koshibae2016theory} and films with the interfacial DMI and perpendicular magnetic anisotropy (PMA), \cite{pierobon2018skyrmion} are accompanied by the nucleation of skyrmion-antiskyrmion or meron-antimeron pairs during the reversal process \cite{heo2016switching}. It is also proposed that for increasingly negative magnetic fields, a defect-free triangular SkX will undergo a first-order topological phase transition to an inverted skyrmion phase via a regime of transient anti-skyrmions \cite{pierobon2018skyrmion}. This scenario is analogous to the "topological switching" phenomena in the bubble lattice. In the presence of defects, the transition becomes smeared so that it appears continuous \cite{pierobon2018skyrmion} and thus ascribed as "topological melting".\\

Recently, the $\beta$-Mn-type chiral magnet Co$_9$Zn$_9$Mn$_2$, with Curie temperature $T_c=400$\,K, was shown to exhibit a robust metastable SkX state at room temperature after field-cooling (FC) through the equilibrium SkX phase that is stable just below $T_c$ \cite{karube2017skyrmion}. At room temperature, the practically infinite relaxation time of this metastable state makes this compound very promising for certain applications, and a model system for investigating the topological stability of metastable magnetic skyrmions. Indeed the enhanced stability of the metastable SkX in Co$_9$Zn$_9$Mn$_2$ at room temperature allows for the investigation of the field-induced transformations of the skyrmion spin texture with a reduced influence of the thermal fluctuation effect \cite{wild2017entropy}. Here we exploit the robust metastable room-temperature SkX in thin plate samples of Co$_9$Zn$_9$Mn$_2$ to address the magnetic field-driven magnetization reversal process experimentally, and investigate the "topological switching" or "topological melting" scenarios.
 
\section{Results and discussion}

\begin{figure*}
\begin{center}
\includegraphics[width=1\linewidth]{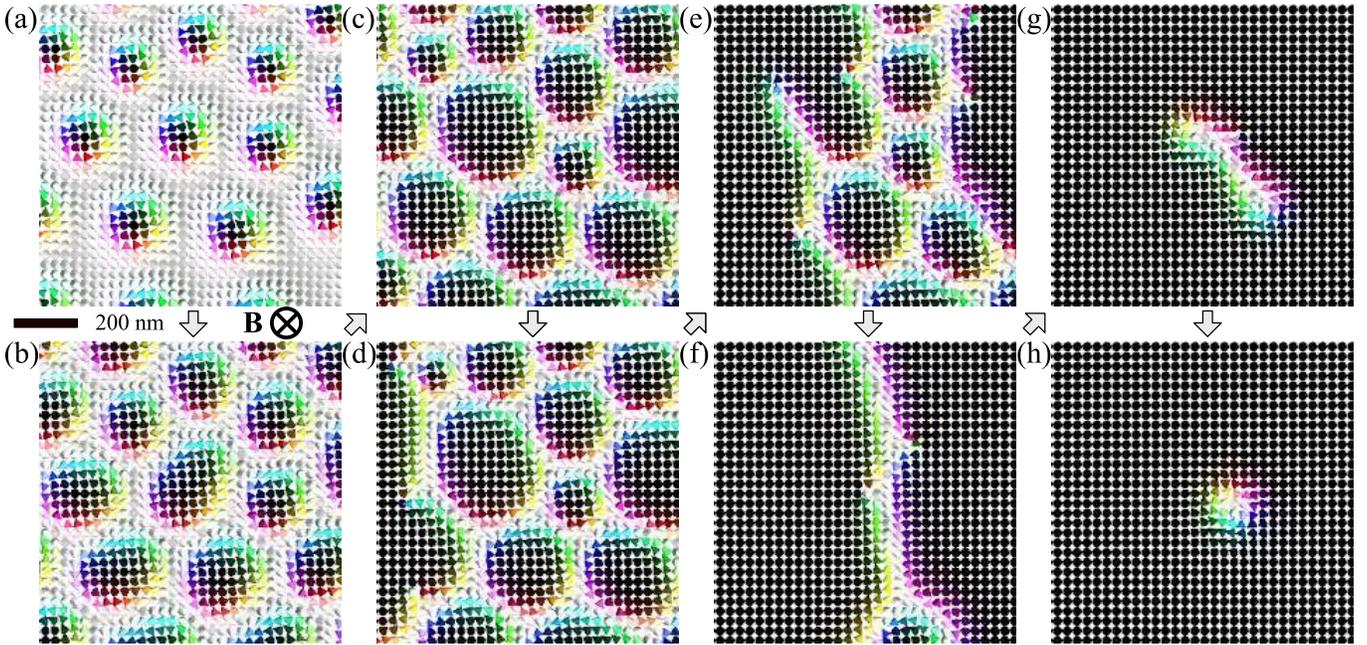}
\caption{Magnified view of the magnetization patterns obtained from micromagnetic simulations in (a) positive field $B=+225$\,mT, (b) zero field, and negative fields (c) $B=-275$\,mT, (d) $B=-285$\,mT, (e) $B=-290$\,mT, (f) $B=-295$\,mT, (g) $B=-300$\,mT, and (h) $B=-310$\,mT.}
\label{fig3}
\end{center}
\end{figure*}

\begin{figure}
\includegraphics[width=1\linewidth]{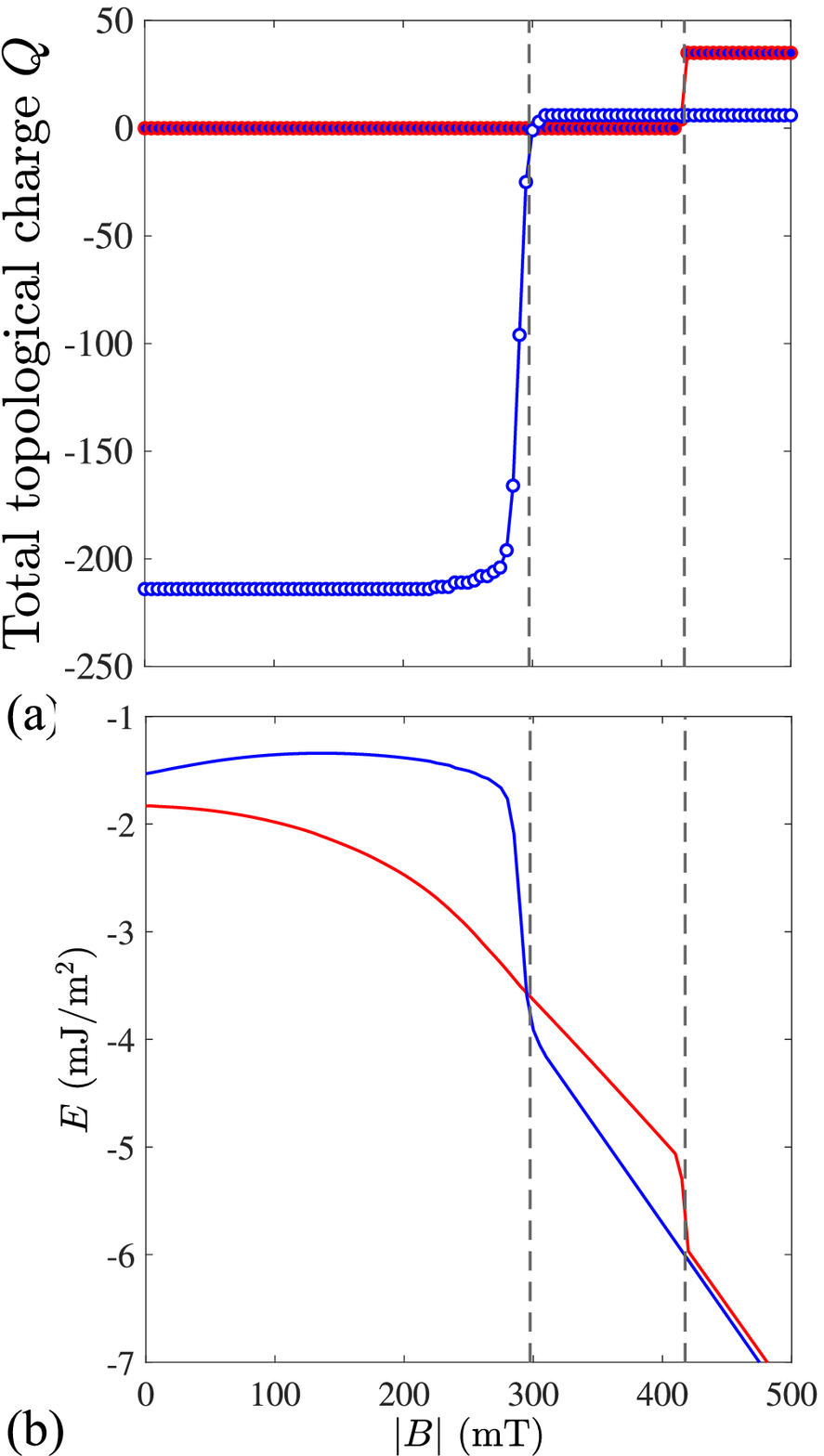}
\caption{Plot of (a) the total topological charge $Q$, and (b) the total energy of the system against external negative field (i.e. a field anti-parallel to the cores of the skyrmions initially stabilised in the case of SkX inversion) derived the micromagnetic simulations. The cases of SkX inversion and SkX nucleation from the helical phase are indicated respectively in each panel by blue and red curves.}
\label{fig4}
\end{figure}

A schematic illustration of both the magnetic phase diagram of  Co$_9$Zn$_9$Mn$_2$ and the measurement protocol are given in Figure \ref{fig1}a. A sketch of the magnetic texture evolution upon magnetic field reversal just below $T_c$ is shown schematically in the middle of the panel. In this case, the thermal equilibrium SkX transitions to the helical structure through the merging of skyrmions firstly into elongated bands, followed by the coherent nucleation of a hexagonal lattice of the reversed skyrmions. This scenario has been confirmed for a 100\,nm-thick plate of FeGe at $T=240$\,K \cite{kovacs2016lorentz}. We note that in the current setup, it was not possible to test the same protocol of SkX reversal in the thermal equilibrium SkX phase of Co$_9$Zn$_9$Mn$_2$ at $T\sim370$\,K, \cite{karube2016robust} since the magnetic field direction provided by the objective lens of the TEM could not be reversed. Therefore, the metastability and significantly long lifetime of the room-temperature and zero-field skyrmions in Co$_9$Zn$_9$Mn$_2$ was of a critical importance.

The main experimental finding of the present work is shown in Figures \ref{fig1}b--k. Initially, a triangular-lattice SkX state with a lattice constant $a_{Sk}\approx150$\,nm is formed as the equilibrium state at approximately 370\,K and 70\,mT \cite{karube2017skyrmion}. Upon FC the sample in $B=70$\,mT from the equilibrium SkX phase to room temperature (290\,K), a robust metastable SkX is achieved which serves as our starting state (Figure \ref{fig1}b). As discussed previously for a bulk sample of Co$_9$Zn$_9$Mn$_2$ \cite{karube2017skyrmion}, the metastable SkX at 290\,K persists even after the magnetic field is reduced to zero. We find however, that already at zero field the expansion of skyrmion cores and the merging of some skyrmions into larger structures indicated by dashed blue lines in Figure \ref{fig1}c. These effects lead to a reduction of the overall topological charge (i.e. number of skyrmions). Notably, no transformation from the SkX to either a helical state \cite{kovacs2016lorentz,peng2018relaxation} or meron-antimeron lattice \cite{yu2018transformation} is observed in weak magnetic fields. 

Since the magnetic field direction provided by the objective lens of the TEM cannot be reversed, the investigation of the negative field evolution of the magnetization distribution was done by removing the sample from the TEM at zero field, and turning it upside-down before re-installing it back into the TEM. The defocus distances for the corresponding LTEM images before and after the $180^\circ$ reversal of the sample were reset from positive (+288\,$\mu$m) to negative (-288\,$\mu$m) values (Figure \ref{fig1}d), and show almost the same magnetic contrast at zero field. Note that after the sample rotation, the applied external magnetic field is in the same direction in the laboratory frame of reference, i.e. parallel to the skyrmion core magnetization of the starting metastable state. Hereafter, we denote the original direction of the field during FC as positive, and the field applied after the sample reversal as negative. \\

By increasing the magnitude of the field in the negative direction, skyrmions continue to merge with each other and the total topological charge decreases further (Figure \ref{fig1}e). Upon sweeping the magnetic field from 0 to -50\,mT, the merging of skyrmions results in the creation and inflation of large skyrmion bubbles with a size of $5-10 a_{Sk}$ (Figure \ref{fig1}e). This phenomenon is consistent with a theoretically predicted scenario in which the SkX transforms into highly unstable large hexagonal cells, which eventually elongate into spiral states \cite{mcgrouther2016internal}.\\

On further increment of the negative magnetic field the bubbles expand, and we observe a peculiar web of skyrmions and bubbles connected to each other by $2\pi$ domain wall strings (Figure \ref{fig1}g-h). In this configuration, the total topological charge still has the same sign as the original SkX state, though the magnitude is much smaller. Skyrmion core reversal occurs for negative magnetic fields below -100\,mT (Figure \ref{fig1}i-j). The inversion of the skyrmion core magnetization is accompanied by the reversal of the rotation sense of the in-plane magnetization, which shows up experimentally by a change in LTEM contrast: skyrmions with core magnetizations aligned antiparallel to the magnetic field appear as white dots (indicated by dashed red circles in Figures \ref{fig1}i-j). Interestingly, the reversed skyrmion core is much smaller than the original one at 0\,T, and its profile changes from hexagonal to circular. This can be understood as a field-driven core collapse, similar to a previous observation of the metastable SkX in FeGe under a high positive magnetic field \cite{yu2018aggregation}. This core collapse behavior has been predicted generally to take place in chiral magnets with uniaxial anisotropy, \cite{bogdanov1994thermodynamically} and is also observed in a skyrmion-hosting monolayer system \cite{romming2015field}. Crucially, in the reversed state the total topological charge $Q$ (number of reversed skyrmions) has changed sign, while the magnitude evaluated for the presently studied viewing region of the sample is $\sim1\%$ of its original value. Finally, all skyrmions are destroyed by a magnetic field of $B=-461$\,mT and the sample crosses over to a conical or induced ferromagnetic state (Figure \ref{fig1}k).\\

An interesting question concerns the detailed magnetic structure of the inverted skyrmions and, especially, the $2\pi$ domain walls appearing upon the magnetization reversal. In principle this could be determined by means of a transport-of-intensity equation (TIE) analysis when the timescale of the magnetic dynamics are slower than that of the relevant LTEM measurements. Unfortunately, due to the metastable character of the observed textures, and, consequently, their dynamic behaviour at the timescale of tens of minutes, it was not possible to measure the series of images with different defocus distances required for a TIE analysis. However, this might be possible in future experiments at lower temperatures where the lifetime of metastable states exceeds the time scale of the experiment.

Next we describe how the experimentally observed negative magnetic field-induced collapse of the SkX agrees with the results of Landau-Lifshitz-Gilbert (LLG) simulations. Our calculations were performed using the MuMax$^3$ package \cite{vansteenkiste2014design}, and appropriate magnetic parameters for a Co$_9$Zn$_9$Mn$_2$ thin plate \cite{takagi2017spin}. The details of the micromagnetic simulation and the simulation code are given in the Supplementary materials. The topological charge on a discrete lattice was quantified using the definition of Berg and L{\"u}scher \cite{berg1981definition}.\\

To follow the experimental FC protocol, we started the simulation with a random initial magnetization and applied a magnetic field in the positive direction. After 2\,ns of relaxation, the magnetic field was removed and the resultant zero-field SkX had a negative field imposed. The magnetization of the system in negative magnetic fields was recorded every 5\,mT with a relaxation time of 2\,ns between the steps. The results are shown in Figure \ref{fig2}. Although the magnitudes of the magnetic fields in the simulation differ from the experimental values, the process of the skyrmion collapse is qualitatively reproduced. The discrepancy between the measured and simulated negative magnetic field ranges of skyrmion stability can originate from thermal fluctuations, the presence of defect sites, or the role of demagnetization fields in the real experiment. \\

Consistent with our real-space observations, the simulations show that the collapse process of the SkX starts from elongated (merged) skyrmions that are already present in the sample at zero field (Figures \ref{fig2}a--c). Furthermore, the expansion of skyrmion cores and their merging behavior results in the formation of large skyrmion bubbles (Figure \ref{fig2}c). The annihilation of these bubbles is accompanied by a contraction of $2 \pi$ domain walls that separate the inside part of the bubble from the outside ferromagnetic medium (Figure \ref{fig2}d). These quasi-one-dimensional topological domain walls tend to collapse into small skyrmions with core magnetizations anti-aligned with both the applied field and the original skyrmion core magnetization (Figure \ref{fig2}e). Eventually, the domain walls vanish, and only a low density of core-inverted skyrmions is observed for sufficiently negative magnetic fields (Figure \ref{fig2}f). The in-plane magnetization direction of the skyrmions in Figures \ref{fig2}a--f is given by the colors shown in Figures \ref{fig2}g,h. A magnified schematic showing the reversal of the skyrmion internal structure in more detail is shown in Figures \ref{fig3}a-h. Recent theoretical \cite{buttner2018theory} and experimental \cite{birch2022history} works have identified the large change in skyrmion size under positive magnetic field, and formation of the large patch-like structures for negative magnetic fields, as signatures of predominantly dipolar-stabilised spin textures in magnetic thin films and multilayers with the interfacial DMI. We note that in Co$_9$Zn$_9$Mn$_2$ the spiral pitch and the skyrmion size are independent from the sample dimensions, and show similar periodicities in bulk crystals and a few-hundred nanometer thick plates \cite{karube2016robust}. Therefore, while the dipolar interaction may impact the behaviour of the SkX under positive or negative magnetic fields, the ground state periodicity $\lambda$ is determined by the dominant DM and exchange interactions.\\

For comparison, we applied the same protocol to determine the total topological charge and system energy for equilibrium SkX nucleation from a starting zero-field helical state. In this case, the magnetization evolves from a single-domain helix to helicoidal structure, and finally, to a a SkX pattern upon increasing the field.

The field-dependence of the calculated topological charge for the processes of both SkX inversion and SkX nucleation from helical order are shown in Figure \ref{fig4}a. The field region $300<|B|<420$\,mT over which the SkX inversion takes place is highlighted.. The simulation results for SkX collapse are qualitatively consistent with the LTEM data: the gradual expansion of the metastable skyrmions takes place upon sweeping the field from 0 to $\sim290$\,mT, followed by inversion of the total topological charge $Q$ from negative to positive. Experimentally, we find the magnitude of topological charge reversal to be $~1\%$ which is in good agreement with a reversed skyrmion fraction of 2.8\% suggested by the simulation. Surprisingly, we find SkX nucleation from the helical phase requires a much higher negative magnetic field than that required for the inversion of metastable skyrmions (Figure \ref{fig4}a). A small hump in the magnetic field dependence of the total $Q$ during the nucleation process at $|B|=280-320$\,mT (see Figure S1 in the Supplementary material \cite{supplementary}) is possibly due to the annihilation of meron defects and single helicoids followed by the creation of isolated skyrmions. At higher field, skyrmions emerge coherently from the rest of the helicoidal texture. The absolute values of the final topological charge in the SkX inversion and SkX nucleation scenarios are not the same due to the difference in the initial number of skyrmions in the starting condition (random spin orientation vs helical pattern). Figure \ref{fig4}b shows the magnetic field dependence of the total system energy. In the whole field range from zero to $|B|\approx300$\,mT the helical (helicoidal) state is energetically favorable, thus emphasizing the metastable character of the starting SkX considered for the SkX collapse process. Crucially, we find the system achieves an energy gain through the stabilization of skyrmions with inverted core magnetizations. This underlines the energetic advantage of the direct skyrmion inversion scenario compared to a SkX to helicoid transition under the application of negative magnetic fields. Interestingly, this energy gain is not reproduced in a simulation with of a sample of smaller dimension (see Supplementary material \cite{supplementary}) due to finite-size effects and the formation of skyrmion bags in this case. A similar phenomena was observed in a 150\,nm-thick FeGe thin plate at low temperature, when field-cooled metastable skyrmions were packed into a bundle by a field cycling procedure \cite{tang2021magnetic}.

Overall, our experimental results are consistent with a field-driven, second-order "topological melting" transition of the metastable SkX in the presence of defects described in Ref. \citenum{pierobon2018skyrmion}. In contrast to the "topological switching" phenomena mediated by anti-skyrmions, the melting scenario does not include the conservation of the total topological charge in the system $|Q|=|n|\cdot N_{sk}$, where $n$ is the topological number as defined in the Equation \ref{eq1}, and $N_{sk}$ is the number of skyrmions in the sample. As the total number of skyrmions $N_{sk}$ is reduced in the melting process to $\sim1\%$ of its original value, only partial topological charge inversion takes place (Figure \ref{fig4}a). The described process of the SkX inversion is qualitatively similar to the classical process of magnetic bubble domain array reversal in PMA systems without DMI, also known as "topological melting" \cite{babcock1989topological,zablotskii1995ordering,jagla2004numerical}.

In Ref. \citenum{pierobon2018skyrmion} the presence of defects was found essential to induce skyrmion melting process, where the defect site acts as a "topological-charge sink". Note, that our simulations did not imply any artificially introduced defects, such as randomly distributed non-magnetic sites or regions with pinned magnetization or anisotropy axes. Instead, the distorted SkX as seen in Figure \ref{fig3}a is formed by the relaxation of random magnetization pattern in an applied field (see Supplementary materials for simulation details). The packing defects of this lattice are similar to the ones observed in the experimental data (Figure \ref{fig1})b and assist the "topological melting" scenario.

\section{Conclusions}

Taken together, our LTEM experiments and micromagnetic simulations clearly demonstrate for the first time a successful inversion of the topological charge in a thin plate of a chiral magnet. The negative field-driven transformation of the SkX is conceivably due to the enhanced energetic stability (so-called topological protection) of zero-field skyrmions at room temperature compared with the thermodynamically stable helical phase. This scenario contrasts strongly with the direct skyrmion-to-helical conversion reported for metastable skyrmions in FeGe and Fe$_{0.5}$Co$_{0.5}$Si \cite{kovacs2016lorentz,peng2018relaxation,milde2013unwinding}. Our micromagnetic simulations of both the skyrmion inversion and skyrmion nucleation processes qualitatively explain our results from the point of view of the energetics. The time resolution of present LTEM observation was not sufficient to describe this process dynamically and identify anti-skyrmion creation that possibly accompanies the skyrmion merging process \cite{koshibae2016theory,pierobon2018skyrmion}. This issue should be further addressed by ultrafast pump-probe LTEM and x-ray imaging studies \cite{berruto2018laser,buttner2015dynamics,litzius2017skyrmion}. 
Finally, we have shown that the negative field-driven merging of metastable skyrmions leads to the formation of large skyrmion bubbles separated by $2\pi$ domain walls. The realization of such large skyrmion bubbles in Co$_9$Zn$_9$Mn$_2$ is furthermore a promising step towards the engineering of novel chiral magnetic particles with any integer topological charge ("skyrmion bags" and "skyrmion bundles" ) considered in the theory of Refs. \cite{rybakov2018chiral,foster2018composite}, and recently realized experimentally in FeGe \cite{tang2021magnetic}. In this instance, it is expected that smaller skyrmions can be nucleated inside the large bubbles by electric current \cite{yu2017current}, optical and x-ray light pulses \cite{berruto2018laser,je2018creation,guang2020creating} or a scanning tunneling microscope tip \cite{wieser2017manipulation}, thus providing new and novel means for topological charge modulation and control.
 
\section{Experimental Section}

\subsection*{Samples}
The details of sample preparation and the structure of the magnetic phase diagrams in both bulk and thin plate samples are given in Ref. \citenum{karube2017skyrmion}.

\subsection*{Real-space imaging}
The real-space imaging of the negative magnetic field-driven collapse of the metastable SkX in a $\sim150$\,nm-thick Co$_9$Zn$_9$Mn$_2$ plate was carried out by means of Lorentz transmission electron microscopy (LTEM). Measurements were performed with a transmission electron microscope (JEM-2100F) at an acceleration voltage of 200\,kV, and the defocus distances of $\pm 288$ $\mu$m as described in the main text.

\section*{Supplementary Information} 

Supporting Information is available from the Wiley Online Library at the link \cite{supplementary} or from the author.

\section*{Acknowledgements} \par
This research was supported in part by JSPS Grant-in-Aids for Scientific Research (Grant No. 20K15164). V.U. and J.S.W. acknowledge funding from the SNSF Sinergia CRSII5\textunderscore171003 NanoSkyrmionics.

\end{document}